\documentclass[10pt,twocolumn,prl,aps,showpacs,superscriptaddress,floatfix,preprintnumbers,reprint,nofootinbib]{revtex4-1}
\usepackage{color}
\usepackage{epsfig}
\usepackage{amsmath,bm}
\usepackage{subcaption}
\usepackage{blindtext}
\usepackage{braket}

      \makeatletter
      \newcommand*{\textalltt}{}
      \DeclareRobustCommand*{\textalltt}{%
	      \begingroup
	      \let\do\@makeother
	      \dospecials
	      \catcode`\\=\z@
	      \catcode`\{=\@ne
	      \catcode`\}=\tw@
	      \verbatim@font\@noligs
	      \@vobeyspaces
	      \frenchspacing
	      \@textalltt
      }
      \newcommand*{\@textalltt}[1]{%
	      #1%
	      \endgroup
      }
      \makeatother

\begin{document}

\title{Neutrino Physics in TeV Scale Gravity Theories}

\author{Manuel Ettengruber}\email{manuel@mpp.mpg.de}

\affiliation{Max-Planck-Institut f\"ur Physik
(Werner-Heisenberg-Institut), F\"ohringer Ring 6, 80805 M\"unchen,
Germany}

\date{\today}

\begin{abstract}
In this paper, the general features of the neutrino sector in TeV scale quantum gravity theories, such as ADD and Many Species Theory, is investigated. 
This class of theories has an inherent way to generate small neutrino masses.  After reviewing this mechanism it is   
generalized to a realistic three-flavour case.  Furthermore,
a procedure is presented how to  diagonalize a mass matrix which is generated by this class of theories and how one can find the Standard Model  flavour eigenstates. The developed general approach is applied to two specific scenarios within ADD and Many Species Theory and possible effects on neutrino oscillations and on unitarity of the lepton mixing matrix are calculated. 
Finally, 
a short overview of phenomenology which can be 
potentially testable by the nowadays neutrino experiments is presented.

\end{abstract}

\maketitle

\section{\label{sec:Intro}Introduction}
\onecolumngrid
The Standard Model (SM) of particle physics
 is one of the most successful theories. 
 In particular, it fully accounts for all 
 the processed  data from high energy particle physics 
 accelerators.   
 
 
  Nevertheless, there are several hints that the SM is not complete. 
 In particular it produces two outstanding puzzles that 
 will be the subject of the present paper:   
 1) the origin and inexplicable smallness of the neutrino mass; 
  2) the Hierarchy Problem.



The hierarchy problem is perhaps the most prominent naturalness 
puzzle of the SM.  
Gravity plays a defining role in its essence due to the following \cite{Dvali:2016ovn}. 
The Higgs mass is quadratically sensitive towards the cutoff 
of the theory.  The ultimate cutoff is provided by gravity 
in form of the Planck mass. This cutoff is fully non-perturbative, since the Planck mass is an absolute upper 
boundary on the mass of elementary particles.  
Indeed, any elementary object much heavier than the Planck scale
is a classical black hole. 

This raises the question what keeps the observed value 
of the Higgs mass-term by some 34 orders of magnitude smaller than the expected upper limit. 
  This Hierarchy strongly hints towards some new stabilizing 
  physics not far from the weak scale. 

One mechanism for stabilization is based on 
lowering the fundamental scale of quantum gravity. 
In this framework, the Planck mass $M_P$ still sets the coupling 
strength of graviton at large distances. However, the 
actual scale $M_*$, at which the quantum gravitational effects are strong, is much lower.  Correspondingly, in such a 
scenario the cutoff-sensitive corrections  
to the Higgs mass 
are regulated  by the scale $M_*$ 
and not by $M_P$.  
  This idea was originally proposed in the ADD model 
 \cite{ArkaniHamed:1998rs, PhysRevD.59.086004}. 
  (see, \cite{Antoniadis:1998ig}, for string theoretic realization). 

  In this setup the fundamental 
  scale of gravity is lowered due to a large volume 
  of extra dimensions.  The reason is that, due to universal nature 
  of gravity, the graviton wavefunction spreads over the entire volume 
  of extra space and gets effectively ``dilute". 
  As a result, the coupling scale of graviton $M_P$
 is hierarchically larger than the fundamental scale 
 of quantum gravity $M_*$. 
    
 Remarkably, more recently it was shown 
 \cite{Dvali:2007hz, Dvali:2007wp} (for various aspects, see, \cite{Dvali:2008fd, 
 Dvali:2008ec, Dvali:2009ne, Dvali:2010vm, Dvali:2021bsy})
 that  the effect of lowering the cutoff  $M_*$ relative to Planck mass $M_P$ is an universal property of any theory with large number
 of particle species.  Correspondingly, a general solution 
 to the hierarchy problem based on this mechanism was proposed in \cite{Dvali:2007hz} and further studied is subsequent papers. 
 We shall refer to this as the ``Many Species" framework.  
 
 As explained in \cite{Dvali:2007hz}, 
 the  ADD model of large extra dimensions represents a particular 
manifestation of this very general phenomenon. There  
the role of the species is assumed by the Kaluza-Klein (KK) 
excitations of graviton.   
This connection enables to understand the dilution 
of graviton wavefunction in extra space of ADD as a dilution in 
what one can call \cite{Dvali:2008fd}
the ``space of species".  In the latter paper, it was argued that,
due to unitarity and other  general consistency properties,  
the  ``space of species"  in many respects behaves as ordinary geometric space. 

In a particularly interesting realization of ``many species" solution 
 the role of the species is played by the 
identical copies of the SM \cite{Dvali:2007hz, Dvali:2007wp}.  
Various phenomenological aspects of this proposal were studied 
in \cite{Dvali:2009ne}.

 Soon after  the invention of low scale quantum gravity idea, 
it has been realized, first in \cite{Arkani-Hamed:1998wuz, Dvali:1999cn} within ADD and later in  
\cite{Dvali:2009ne} within many species theory, that this   
general framework, as a bonus, offers an universal solution  
to the neutrino mass problem in the SM. 
Namely, the same mechanism that explains the hierarchy 
between the weak and Planck scales, is responsible for 
the suppression of the neutrino mass. 
 
\par

Naturally, this fact boosts the motivation for the above class of theories, 
since the origin and the hierarchy of the neutrino mass is 
a fundamental open question in SM. 
Due to the phenomenon of neutrino oscillations \cite{Fantini2018}, we know that they have a mass and this is now a field of several 
ongoing experiments. 
So far,  neutrino mass has not been detected directly.
Just upper bounds of roughly $1$eV have been given \cite{Aker2019}. It is also not known 
 whether the neutrino mass is of Majorana or of Dirac nature as this is the case with all other fermions of the SM.
 We also do not know why neutrinos are so much lighter
than the charged fermions. 

 If neutrino masses are of Dirac type and originate from
 an ordinary Higgs mechanism, an unusually-small 
Yukawa  coupling would be required 
for generating the masses smaller than eV.  
 
Lack of the explanation for this smallness, 
prompted thinking that perhaps the neutrino mass
is of Majorana type. In such a case, the 
mass term must be generated from an effective 
high-dimensional operator \cite{Weinberg:1979sa} 
and the smallness 
can be attributed to the high cutoff scale.  
In the traditional See-Saw mechanism
\cite{Minkowski:1977sc, GellMann:1980vs, Yanagida:1980xy, Mohapatra:1979ia, Mohapatra:2004zh},  such an 
effective operator is generated by integrating out 
a neutrino's hypothetical right-handed partner with large Majorana mass. \\

 However, after formulation of theories with low scale 
 gravity, it became clear 
 that they offer 
 an alternative possibility in form of naturally small Dirac 
neutrino mass.  Originally,  this  idea was realized in 
\cite{Arkani-Hamed:1998wuz, Dvali:1999cn} within 
the ADD framework \cite{ArkaniHamed:1998rs}. We shall refer to this as ADDM model. 
Later a complementary mechanism of suppressed neutrino mass 
was introduced by Dvali-Redi (DR) in \cite{Dvali:2009ne} within the 
``many species" framework with identical copies of the SM
\footnote{Such scenarios have other potential bonuses. For example, 
particles of other copies could just interact with each other gravitationally and are good candidates for dark matter \cite{Dvali2009b, Dvali:2009ne}}.  
 
 Although complementary,  the above two scenarios are based 
 on one and the same fundamental mechanism of the 
suppression of the neutrino Yukawa coupling, very similar 
to the suppression of the coupling of graviton. 
In both cases this can be viewed as a consequence of the 
dilution of the wavefunction of the sterile neutrino 
into the bulk of the extra space of species. 
This dilution is identical to the dilution of the wavefunction 
of graviton in the same space.  
In ADD this space is also organized as a real coordinate space but 
this does not change the essence of the dilution. 
  In summary, the theories with low $M_*$ can solve both the hierarchy and the neutrino mass problems by one and the same mechanism.   
Both hierarchies are controlled by the ratio
$M_*/M_P$.

The main focus of the present work is implications for neutrino physics.  The above class of theories predict certain universal features of phenomenological interest.  In particular 
as already discussed in  
\cite{Dvali:1999cn} for ADDM  and in \cite{Dvali:2009ne} for DR 
scenarios, the mixing with the tower of sterile neutrino species results into 
oscillations of neutrinos into the hidden sector. The implies  
non-conservation of neutrino number within SM and correspondingly 
a seeming violation of unitarity.  
 This is of obvious potential experimental interest.

 The aim of this paper is to focus on the neutrino sector of these theories and work out a general framework for how neutrino physics can be treated in this class of theories. Moreover, we will generalize this framework to a realistic three flavour case and investigate their effects on low energy phenomena  and observables  such as neutrino oscillations into the hidden modes and possible deviations from the Standard Model PMNS matrix which can be tested by nowadays experiments. This is a still ongoing project \cite{project}. \\

This paper is organized as follows. In section 2 the structures of DR and ADDM models are presented and reviewed. In section 3 we formulate a general approach how neutrino masses are induced in this kind of theories and how we can guarantee the smallness of their mass. In section 4 we present a generalisation of the mass matrices to a realistic three flavour case and in section 5 we investigate the special case of highly symmetric mass matrices which are of interest for the DR model. In section 6 the phenomenology of TeV scale gravity theories in neutrino physics are investigated. In section 7 we wrap up our findings and give an outlook to possible experimental tests.

\section{\label{sec:ParamSpaceLikePriors}ADD and "Many Species Theory"}
\subsection{Many Species Theory}
 Originally the idea of many mirror copies of the SM has been 
proposed in \cite{Dvali:2007hz, Dvali:2007wp} as the framework 
for solving  the hierarchy problem. 
In this work, it has been shown 
that through introducing  $N$ particle species, the fundamental scale of gravity $M_{*}$ in lowered relative 
to Planck scale $M_P$
 in the following way 
\begin{equation} \label{Ncentral}
M_p^2 \geq N M_{*}^2.
\end{equation}
This result has been obtained using the well established properties of 
black hole physics and is fully non-perturbative.  
In order to lower the scale of gravity to TeV energies, $N$ must 
be of the order of $10^{32}$.  Since the bound (\ref{Ncentral}) 
is independent of the nature of the particles,  the species of 
various types can be used 
 for lowering the gravitational cutoff.  
 A particular version introduced in  \cite{Dvali:2007hz, Dvali:2007wp}, assumes that the species are identical copies of the SM. 
 Phenomenological; aspects including generation of neutrino 
 mass were discussed in  \cite{Dvali:2009ne}.
 \par

In this work, it is assumed that the species obey the full permutation group $P(N)$ at first. 
 This means that all species are equidistant in 
what one can call the ``space of species".  
This gives a certain predictive power to the theory. 
Alternative choices such as the cyclic symmetry are also possible. 
  It was shown that many species framework can give 
  various phenomenological signatures, including micro black holes, 
  in region of TeV energies. In the present work we shall focus on 
  the implications for neutrino masses. \par  


 We shall study generalizations of the mechanism, 
originally introduced in  \cite{Dvali:2009ne},   
 which allows the generation of small neutrino masses in many species framework. This mechanism represents an
 infrared alternative to see-saw which cannot be used 
 in frameworks with low cutoff.  
 

Let us briefly review the results of \cite{Dvali:2009ne}.
 As already said, the framework represents $N$ identical copies of the SM.  
The copies are permuted under $P(N)$.  It is useful to 
visualize the copies as placed on equidistant sites in the 
space of species.  Fermions are each sector is charged under their  own gauge group. The exceptions are sterile neutrinos, which 
represent the right-handed partners of corresponding 
active left-handed neutrinos.   We shall denote them by 
$\nu_{Rj}$, where $j =1,2,...,N$ is the label of the SM copy. 
These  particles do not carry any charges under the SM
gauge groups. Thus the notion of ``belonging" is defined by 
their transformation properties under the permutation group 
$P(N)$ as well as by their couplings to particles of specific  SM 
copies. 
In particular, the gauge charges  do not forbid sterile neutrinos 
to interact with neutrinos of the other copies. 
One can say that sterile neutrinos are not confined  
to specific sites in the space of species.  
The most generic renormalizable coupling has the following
structure,  
\begin{equation}
(HL)_i \lambda_{ij}\nu_{Rj} + h.c. ,
\end{equation}
where $H$ and $L$ stand for the Higgs and lepton doublets 
of the $i$-th copy.  
Here $\lambda_{ij}$ is a $N \times N$ Yukawa matrix interaction in the space of species. This Yukawa coupling matrix is restricted by the permutation symmetry group $P(N)$ and has therefore the following form: 

\begin{equation}
 \lambda_{ij}=\begin{pmatrix}
 a&b&b&\dots\\
 b&a&b&\dots\\
 b&b&a&\dots\\
 \dots&\dots&\dots&\ddots
\end{pmatrix}. 
\label{Yukawa}
\end{equation}
For calculation of the mass matrix of neutrinos, one has to have a closer look at the Higgs doublet $H_i$. 
The simplest case for calculation is when the permutation 
symmetry is unbroken by the electroweak vacua. 
That is,  VEV of the Higgs doublet  in every copy of the SM takes the same value $v$. In this section we shall focus on this case. 
The generalization to the case of broken permutation symmetry 
will be given later.  

 
For now,  let us therefore take $v$ as the VEV of the Higgses for all copies.  Then,  the mass matrix takes the form $ m_{ij}=\lambda_{ij} v$. \par

This mass matrix has the eigenvalues
\begin{equation}
m_1^{\prime} = (a-b)v,
\end{equation}
\begin{equation}
m_H = [a+(N-1)b]v,
\label{mH}
\end{equation}

corresponding to the eigenvectors

\begin{equation}
\nu_1^{\prime}= \sqrt{\frac{N-1}{N}}\nu_1-\frac{1}{\sqrt{N}}\nu_h,
\label{eig1}
\end{equation}
and 
\begin{equation}
\nu_H = \frac{1}{\sqrt{N}}\nu_1 + \sqrt{\frac{N-1}{N}}\nu_h.
\label{eigH}
\end{equation}
It is worth noticing for later convenience that the light eigenvalue is $N-1$ times degenerated. Because
\begin{equation}
    b \leq \frac{1}{\sqrt{N}},
\end{equation}
and $a \approx 100b$ we see that the mass of the neutrino is suppressed by the number of species. The mechanism presented here can explain the smallness of the neutrino mass but has no phenomenological implications which can be tested by experiments due to the huge mass of the heavy state which scales with the number of species which is of order $N\approx 10^{32}$.

\subsection{The ADDM model}
The ADDM model \cite{ArkaniHamed:1998rs, Antoniadis:1998ig,  PhysRevD.59.086004} is based on the idea that in addition to 
observed $3$ space dimensions, there exist $d$ additional compact space  
ones with radii $R_i, ~i=1,2,...,d$ below the tenths of a millimetre.

The role of the gravitational cutoff in this theory is played 
by the fundamental Planck mass of the $4+d$-dimensional theory, $M_f$. The two Planck scales are related via, 
\begin{equation}
M_P = M_f \sqrt{M_f^d V_d},
\label{ADDcutoff}
\end{equation}
where 
\begin{equation}
V_d = (2\pi)^d R_1 ... R_d,
\end{equation}
is the volume of the extra dimensional space. 

  Like the DR, this theory provides a solution to the hierarchy problem by lowering the cutoff $M_f$  relative 
  to Planck mass due to large volume of the extra space. 
  $M_f \sim $TeV requires that the volume of the extra space, measured 
  in units of the fundamental Planck mass, be about  
$M_f^d V_d \sim 10^{32}$.   

 As noticed in \cite{Dvali:2007hz} the lowering of the cutoff 
 in  ADD can be understood as a particular case of many species
 effect.  This is because  the quantity 
 $M_f^d V_d$ measures the number of Kaluza-Klein species 
 of graviton.  Thus, the relation (\ref{ADDcutoff}) 
 represents a particular manifestation of a more general 
 relation (\ref{Ncentral}).


 \par

 According to this theory, the Standard Model particles are localized on a 3-dimensional hyper-surface (brane) which is embedded in the bulk of $d$ large extra dimensions. The graviton propagates into the entire high dimensional space.  Together with gravity, the bulk is a natural habitat for all possible particles that carry no gauge quantum numbers 
 under the Standard model group.  
 
 Notice that \cite{ArkaniHamed:1998rs}  
 the bulk particles cannot carry any quantum numbers 
 under the SM gauge group.  This is a consistency requirement 
 that follows from the gauge invariance and is an intrinsic 
 feature of localization mechanism of the gauge field on 
 the brane \cite{Dvali:1996xe}. Correspondingly, 
 the localization of SM gauge fields on the brane automaticaly
 forbids existence of any bulk modes with such charges. 
 Only the particles carrying no SM gauge quantum numbers 
 are permitted to represent bulk modes. 
 In particular, such are 
 sterile neutrinos that play the role of the right-handed partners 
 of the ordinary left-handed neutrinos of the Standard Model. 
  
  This setup generates a naturally small  Dirac mass for neutrinos   \cite{Arkani-Hamed:1998wuz, Dvali:1999cn}. This mass originates  
from the mixing of the right-handed component of bulk sterile neutrino $\nu$ with the 
 Standard Model left-handed neutrino $\nu_L$
  which is localized 
  on the brane. 
  In the approximation of a zero-width brane, the 
 part of the action responsible for this mixing can be written as 
   \begin{equation}
\int   d^4x  \frac{h}{M_f^{d/2}}  H(x_{\mu} )\bar{\nu}_L(x_{\mu})
\nu (x_{\mu}, y_i=0) + h.c.,
\label{Coupling}
\end{equation}
 where $x_{\mu}$ stands for ordinary $4$-dimensional space-time coordinates and $y_i,~i=1,2,...,d$-s are the extra ones.  The brane location is taken at 
 $y_i=0$ point. The canonically normalized $4+d$-dimensional fermion field $\nu(x,y)$  has dimensionality 
 $(3+d)/2$.  Correspondingly, the coupling constant has 
 dimensionality $-d/2$.  We have parameterized this coupling constant in terms of the fundamental scale $M_f$ and an order-one  dimensionless constant $h$.   
 
 From the point of view of $4$-dimensional theory, 
 $\nu(x,y)$ represents a tower of 
  Kaluza-Klein modes with their masses quatized 
  in units of the inverse radii  $m^2 = \sum_in_i^2/R_i^2$ 
  where $n_i$ are integers.  
   
   Notice that, a high dimensional 
 fermion field $\nu$, viewed from the point of view of 
 a $4$-dimensional theory, has no chirality. 
 That is, at each Kaluza-Klein level of mass 
 $m$ it contains $4$-dimensional fermions of both 
 chiralities, $\nu_R^{(m)}$ and $\nu_L^{(m)}$.   
 The $4$-dimensional reduction of the coupling 
(\ref{Coupling}) gives   
\begin{equation} \label{redCoupling} 
    \frac{h M_f}{M_P}H \bar{\nu}_{L} \sum_{m}\nu_{R}^{(m)} + h.c. \,,
\end{equation}
where the factor $1/M_P$ comes from the canonical normalization 
of the Kaluza-Klein modes. Notice that only 
the right-handed components $\nu_{R}^{(m)}$ of the Kaluza-Klein modes mix with SM neutrino.

After taking into account the VEV of the Higgs field, $\langle H \rangle \equiv v$, the above couplings 
translate as the Dirac-type mass terms
\begin{equation} \label{DirMix} 
    m_D \bar{\nu}_{L} \sum_{m}\nu_{R}^{(m)} + h.c. \,,
\end{equation}
with $m_D \equiv \frac{h v M_f}{M_P}$.  This mixing generates  
a Dirac mass of the SM neutrino. 
Below, for evaluating the  mass matrix we shall restrict ourselves to the case of a single extra 
dimension. In this case the masses of Kaluza-Klein excitations 
are labeled by a single integer $m = n/R$. 

 Taking into account the  Dirac mass terms of Kaluza-Klein modes
 coming from the mixing between their left and right-handed components,  

\begin{equation}
    \sum_{n= -\infty}^{\infty}\frac{n}{R} \bar{\nu}_{nR}\bar{\nu}_{nL} \,,
\end{equation}
the resulting mass matrix has the form:
\begin{equation}
    M =\begin{pmatrix}
    m_D & \sqrt{2} m_D &  \sqrt{2} m_D & \dots &  \sqrt{2} m_D \\
    0 & \frac{1}{R} &0&\dots&0 \\
    0&0&\frac{2}{R}&\dots&0\\
    \dots&\dots&\dots&\dots&\dots\\
    0&0&0& \dots& \frac{k}{R}
    \end{pmatrix}
\end{equation}
After the diagonalization of the mass matrix, one can express a neutrino of a specific flavour with the following expression
\begin{equation}
\nu = \frac{1}{\Omega}(\nu_0 + \xi \sum_{n=1} \frac{1}{n}\tilde{\nu}_n),
\label{masterADD}
\end{equation}
with
\begin{equation}
\xi = \frac{\sqrt{2}vM_fR h}{M_P}.
\end{equation}
The normalisation parameter is $\Omega^2 = 1+ \frac{\pi^2}{6}\xi^2$. 


The mass of the lightest  eigenstate $\nu_0$ is
\begin{equation}
m_D = \frac{hvM_f}{M_P}.
\label{ADDmassground}
\end{equation}
The other eigenstates have
\begin{equation}
m_n \approx \frac{n}{R}.
\label{ADDmassheavy}
\end{equation}
One of the important phenomenological implications of this scenario is the oscillation of active neutrino species into the KK neutrinos \cite{Dvali:1999cn}.  The effect takes place already for a single flavor case. We shall review this later and compare it with  the case of three flavors of active neutrino species.

\section{\label{sec:Results}Generalisation of Neutrino Masses}
We have seen that in ADDM and in DR one can generate small neutrino masses by introducing a sterile 
neutrino which is uncharged under the SM gauge group and can therefore propagate into an additional space which was introduced in this class of theories. In the case of ADDM, this space is represented the bulk of  large extra dimensions and in the DR it is described as the "space of species".
In both cases, the neutrino mass is suppressed by the large effective volume of this  extra space.

This common structure we want to investigate further.
We shall make a rather general assumption of  the existence of an extra space in which the sterile neutrino can propagate.  
Also, we assume that this extra space lowers the scale of gravity via
\begin{equation}
    M_*^2 = \frac{M_P^2}{\Lambda},
    \label{lowering}
\end{equation}
where $\Lambda$ is the volume of the extra space measured in 
fundamental units. In ADDM the size of the extra space is a function of $R$ $\Lambda(R)$ and in DR of $N$ $\Lambda(N)$.
 
  It is assumed that the particles that are not 
  charged under the SM gauge symmetries can propagate 
  in this extra space.  That is, the couplings of such 
  particles are more or less uniformly spread over this space. 
  Correspondingly, the coupling to individual copies is suppressed.  
  
  Within the known framework there are two candidates for such particles. First one is of course graviton, since gravity interacts 
  universally.  The second natural candidate is a 
  sterile neutrino.  Currently, it is not known whether neutrino is 
  a purely Majorana particle.  If it is not, then there necessarily 
  exist a sterile partner $\nu_R$ that together with ordinary 
 left-handed neutrino forms a Dirac state.  
  This sterile neutrino carries no gauge quantum numbers 
  under the standard model group. Correspondingly, it 
  has no obligation to be confined to the site where our Standard Model is located. Instead, just like gravity, such particles can 
  spread over the entire extra space, irrespectively whether 
 this space stands for extra space dimensions or the space of species. This spread naturally suppressed the coupling 
 of the sterile fermion to SM neutrino, thereby resulting into 
 small Dirac mass. The suppression of the coupling with many mixing partners results from the principle of unitarity and was shown in \cite{Dvali:2008fd}.  This is the key mechanism behind the 
 small neutrino mass both in ADDM \cite{Arkani-Hamed:1998wuz}
  as well as in DR  \cite{Dvali:2009ne}. 

%
  
%


A possible operator for neutrino mass of the SM neutrino is the Dirac operator
\begin{equation}
yH\bar{\nu}_L\nu_R,
\label{Dirac1}
\end{equation}
where $y$ is a Yukawa coupling and $H$ is the SM Higgs doublet. In this framework, the left-handed neutrinos of the SM can mix with different types of right-handed neutrinos which are inhabitants of the extra space. So $\nu_R$ is a superposition of all possible mixing partners
\begin{equation}
\nu_R = \frac{1}{\Lambda} \sum_n c_n \nu_{nR} \,.
\end{equation}
Of course, the superposition has to be normalized and this depends on the size of the extra space the right-handed neutrinos live in. Therefore the different contributions of all mixing partners have to be divided by the volume of space in which they can propagate. The resulting form of (\ref{Dirac1}) is then

\begin{equation}
yH\bar{\nu}_L\nu_R = \frac{yv}{\Lambda} \bar{\nu}_L\sum_n c_n   \nu_{nR}\,.
\end{equation}
With (\ref{lowering}) one gets 
\begin{equation}
\frac{M_f}{M_P}yv\bar{\nu}_L\sum_n c_n   \nu_{nR}.
\label{mixingoperator}
\end{equation}
 The factor in front of the operator represents the effective Dirac mass of neutrino which we can denote by
\begin{equation}
m_D = \frac{M_f}{M_P}yv \,.
\end{equation}
Here we want to point out that this prefactor is suppressed by the Planck mass.
 We see that it induces a small Dirac mass for neutrinos. 
This captures an universal essence of generating a small neutrino 
mass in ADDM \cite{Arkani-Hamed:1998wuz} and in DR 
\cite{Dvali:2009ne}  formulated in a theory-independent way. 
 It follows that a small mass for neutrinos is a natural property for this class of theories. The new feature is that the suppression of the mass of the neutrino comes from the size of the extra space
 to which the sterile neutrino can propagate. 
 This is very different from the introduction of a heavy Majorana 
 particle as this is the case in see-saw.  
  In other words, the spirit of the solution for the smallness of the neutrino mass we presented here is an infrared solution and not an ultraviolet solution by introducing a very heavy particle.\par 

Of course,  such a mixing can also occur between  $\nu_{Rj}$
and the left-handed inhabitants,  $\nu_{Li}$, of the 
extra space. Therefore, we also include the mass terms 
of the following form, 
\begin{equation}
m_{ij}\bar{\nu}_{Li} \nu_{Rj}.
\end{equation}

Let us label the neutrino of the SM with $i= 1$ and redefine the Yukawa coupling as $y=yc_1$. Moreover, let us assume that the interactions among  certain pairs  of neutrinos  are stronger than the mixing with other types. We shall organize such mass terms 
as the diagonal entries $m_{ii}$. Correspondingly the off-diagonal entries $\mu_{ij}$ will denote mixings with other species.
 The resulting mass matrix is

\begin{equation}
\begin{pmatrix}
m_D& \mu_{12}&\dots&\dots\\
\mu_{21}&m_{22}&\mu_{23}&\dots\\
\vdots&\dots&\ddots&\dots
\label{mass}
\end{pmatrix},
\end{equation}
with $\mu_{1i} = c_i m_D$, and we ordered the diagonal entries according to their hierarchy 
\begin{equation}
m_D < m_{22}< \dots < m_{kk}. 
\end{equation}

Assuming that the mixing angles due to off-diagonal entries 
are small, we can split this matrix into the diagonal and off-diagonal parts and treat the latter one as a perturbation 
\begin{equation}
\begin{pmatrix}
m_D& \mu_{12}&\dots&\dots\\
\mu_{21}&m_{22}&\mu_{23}&\dots\\
\vdots&\dots&\ddots&\dots
\end{pmatrix}=\begin{pmatrix}
m_D& 0&\dots&\dots\\
0&m_{22}&0&\dots\\
\vdots&\dots&\ddots&\dots
\end{pmatrix}+\begin{pmatrix}
0& \mu_{12}&\dots&\dots\\
\mu_{21}&0&\mu_{23}&\dots\\
\vdots&\dots&\ddots&\dots
\end{pmatrix},
\end{equation}

and we denote
\begin{equation}
V\equiv \begin{pmatrix}
0& \mu_{12}&\dots&\dots\\
\mu_{21}&0&\mu_{23}&\dots\\
\vdots&\dots&\ddots&\dots
\end{pmatrix}.
\end{equation}
With this, we find that the eigenvalues do not become corrected in  the first order in mixing 
\begin{equation}
m_i = m_{ii} + \braket{n_i|V|n_i} = m_{ii} + \mathcal{O}^2.
\end{equation}
The correction to the mass eigenstates 
has the following form

\begin{equation}
\ket{m_1}= \ket{1^{(0)}}+ \sum_{k=2}\frac{\mu_{1k}}{m_1^{(0)}-m_k^{(0)}}\ket{k^{(0)}},
\end{equation}
where the $\ket{n}$ are the eigenstates of the unperturbed matrix. 
 Of course one has to normalise the expression with
\begin{equation}
Norm^2= 1+ \sum_{k \neq n} (\frac{\mu_{nk}}{m_n^{(0)}-m_k^{(0)}})^2.
\end{equation}

This leads then to the following expression for the mass eigenstates 
\begin{equation}
\ket{\vec{m}}= \begin{pmatrix}
1 & \frac{\mu_{12}}{m_1-m_2}&\dots \\
\frac{\mu_{21}}{m_2-m_1}&1&\dots\\
\vdots&\dots&\ddots
\end{pmatrix} \ket{\vec{n}},
\end{equation}
symbolically
\begin{equation}
\ket{\vec{m}} = U \ket{\vec{n}}.
\end{equation}
Now one has to invert U in order to find the expression for the space states. In order to invert the matrix U we use the equation
\begin{equation}
(A+X)^{-1} = A^{-1} + Y ,
\end{equation}
with
\begin{equation}
Y = -A^{-1}XA^{-1},
\end{equation}
and X beeing in this case the perturbation matrix V. One therefore gets for $U^{-1}$
\begin{equation}
U^{-1} = \begin{pmatrix}
1 & -\frac{\mu_{12}}{m_1-m_2}&\dots \\
-\frac{\mu_{21}}{m_2-m_1}&1&\dots\\
\vdots&\dots&\ddots
\end{pmatrix}.
\end{equation}
 This is how the mixing  with the states of extra space takes place in case of a single flavour of SM neutrino. 
 In particular, the above reproduces the results of
 such mixings in ADDM 
 \cite{Arkani-Hamed:1998wuz} and in DR
\cite{Dvali:2009ne} for the case of a single flavor. 

%

\section{\label{sec:Conclusions}Generalisation to three flavour case}
We now generalize the discussion for the case of 
 three flavors of SM neutrinos. 
The simplest (but unrealistic) case is if all the three flavor neutrinos have their own mixing partners in the extra space. 
In such a case the mass matrix has the following  block-diagonal form

\begin{equation}
\mathcal{M} = \begin{pmatrix}
M_e & 0&0\\
0& M_{\mu}&0\\
0&0&M_{\tau}
\end{pmatrix},
\end{equation}
where the $M_{\alpha}$ stand for the mass matrices of the different flavours. Each of them has a form analogous to (\ref{mass}).
 Of course, we must take mixing among the different flavours into account. This is necessary for phenomenological consistency. In particular,  to make the SM three flavour neutrino oscillations possible. In order to incorporate this phenomenon 
 we have to depart from the above block-diagonal structure. 
  We therefore write, 
\begin{equation}
\mathcal{M}=\begin{pmatrix}
M_e & e\mu&e\tau\\
e\mu& M_{\mu}&\mu\tau\\
e\tau&\mu \tau&M_{\tau}
\end{pmatrix},
\label{MassMatrix}
\end{equation}
where we denote with the $\alpha \beta$ ($\alpha, \beta = e, \mu, \tau)$ entries the mixing matrices among the different space state partners of different flavours. 
In order to increase the precision of the perturbative calculation we treat the mixing of the flavour ground states (i.e., the 
direct mixing among SM neutrinos)  as part of the perturbed matrix and not as a part of the perturbation matrix V. This leads to the following expressions for the mass eigenstates of the three active neutrinos  (we denoted the entries of the SM like mixing elements as $U_{ei}^{-1}$) 

\begin{multline}
\ket{m_1^e}= U_{e1}^{-1}\ket{e}+U_{e2}^{-1}\ket{\mu}+ U_{e3}^{-1}\ket{\tau}+\sum_{k=2}\frac{U_{e1}^{-1}\mu_{1k}^e+U_{e2}^{-1}e\mu_{1k}+U_{e3}^{-1}e\tau_{1k}}{m_1^e-m_k^e}\ket{k_1^e}+\\\\ \sum_{k=2}\frac{U_{e1}^{-1}e\mu_{1k}+U_{e2}^{-1}\mu_{1k}^{\mu}+U_{e3}^{-1}\mu\tau_{1k}
}{m_1^e-m_k^{\mu}}\ket{k_1^{\mu}}+ \sum_{k=2}\frac{U_{e1}^{-1}e\tau_{1k}+U_{e2}^{-1}\mu\tau_{1k}+U_{e3}^{-1}\mu_{1k}^{\tau}}{m_1^e-m_k^{\tau}}\ket{k_1^{\tau}}.
\end{multline}

This is the expression for the lightest mass eigenstate and we identify it with the dominant mass eigenstate for the electron neutrino. We have to invert this expression now in an analogue way as in the one flavour case and in order to do so we assume that 
\begin{equation}
U_{e1} >> U_{e2}, U_{e3} >> e\mu_{1i}, e\tau_{1i}.
\label{nastyassumption}
\end{equation}
Then we can write the interaction eigenstate approximately as

\begin{multline}
\ket{\nu_e}= U_{e1}\ket{m_1^e}+U_{e2}\ket{m_1^{\mu}}+U_{e3}\ket{m_1^{\tau}}- U_{e1}(\sum_{k=2}\frac{U_{e1}^{-1}\mu_{1k}^e+U_{e2}^{-1}e\mu_{1k}+U_{e3}^{-1}e\tau_{1k}}{m_1^e-m_k^e} \ket{m_k^e}\\\\ + \sum_{k=2}\frac{U_{e1}^{-1}e\mu_{1k}+U_{e2}^{-1}\mu_{1k}^{\mu}+U_{e3}^{-1}\mu\tau_{1k}}{m_1^e-m_k^{\mu}}\ket{m_k^{\mu}}+\sum_{k=2}\frac{U_{e1}^{-1}e\tau_{1k}+U_{e2}^{-1}\mu\tau_{1k}+U_{e3}^{-1}\mu_{1k}^{\tau}}{m_1^e-m_k^{\tau}}\ket{m_k^{\tau}}).
\label{generalOscillation}
\end{multline}

  If we assume that $U_{ei}$ are already normalized, 
 the normalisation looks as follows,

\begin{multline}
N_e^2= 1+U_{e1}(\sum_{k=2}\frac{U_{e1}^{-1}\mu_{1k}^e+U_{e2}^{-1}e\mu_{1k}+U_{e3}^{-1}e\tau_{1k}}{m_1^e-m_k^e})^2+ (\sum_{k=2}\frac{U_{e1}^{-1}e\mu_{1k}+U_{e2}^{-1}\mu_{1k}^{\mu}+U_{e3}^{-1}\mu\tau_{1k}}{m_1^e-m_k^{\mu}})^2+ \\\\ (\sum_{k=2}\frac{U_{e1}^{-1}e\tau_{1k}+U_{e2}^{-1}\mu\tau_{1k}+U_{e3}^{-1}\mu_{1k}^{\tau}}{m_1^e-m_k^{\tau}})^2.
\end{multline}

We can simplify the expression for the flavour neutrino a little bit further by assuming that the masses of the bulk states in the diagonal entries are the same for all flavors. This means that
\begin{equation}
    m_k^e = m_k^{\mu} = m_k^{\tau} = m_k.
\end{equation}
 We also want to assume that different cross mixing elements among different flavors have the same structure as the mixing of bulk states with its own flavor. This means that also the mixing parts $\alpha \beta_{1k}$ and $\mu_{1k}^{\alpha}$  look like
 
 \begin{equation}
     \mu_{1k}^{\alpha} = \mu f(m_D^{\alpha}),
 \end{equation}
 with the same overall constant $\mu$ and the same function $f$ depending on the induced Dirac mass just differing by the argument. 
 This leads then to the following expression for the flavor eigenstate
\begin{equation}
  \ket{\nu_e}= U_{e1}\ket{m_1^e}+U_{e2}\ket{m_1^{\mu}}+U_{e3}\ket{m_1^{\tau}}- U_{e1} \sum_{\alpha =1}^3 \sum_{k=1}\frac{\mu_{1k}^e U_{e1}^{-1}+ \mu_{1k}^{\mu}U_{e2}^{-1} + \mu_{1k}^{\tau} U_{e3}^{-1}}{m^e - m_k} \ket{k^{\alpha}}.
  \label{simplifiedgeneralOscillation}
\end{equation}

Now let us drop the assumption (\ref{nastyassumption}) and give for the simplified equation (\ref{simplifiedgeneralOscillation}) the expression for a larger cross mixing among the SM neutrinos which is a more realistic scenario. Then the equation gets modified in the following way
\begin{equation}
    \ket{\nu_e}= U_{e1}\ket{m_1^e}+U_{e2}\ket{m_1^{\mu}}+U_{e3}\ket{m_1^{\tau}}- \sum_{\alpha =1}^3 \sum_{k=1}\frac{\overrightarrow{U_e}\overrightarrow{C}}{m^e - m_k} \ket{k^{\alpha}},
    \label{generalOsc}
\end{equation}

with
\begin{equation}
\overrightarrow{U_e} = \begin{pmatrix}
U_{e1} \\ U_{e2} \\ U_{e3}
\end{pmatrix},
\end{equation}
and
\begin{equation}
   \overrightarrow{C} = \begin{pmatrix}
        \mu_{1k}^e U_{e1}^{-1}+ \mu_{1k}^{\mu}U_{e2}^{-1} + \mu_{1k}^{\tau} U_{e3}^{-1} \\
        \mu_{1k}^e U_{\mu1}^{-1}+ \mu_{1k}^{\mu}U_{\mu2}^{-1} + \mu_{1k}^{\tau} U_{\mu3}^{-1} \\
        \mu_{1k}^e U_{\tau1}^{-1}+ \mu_{1k}^{\mu}U_{\tau2}^{-1} + \mu_{1k}^{\tau} U_{\tau3}^{-1} 
   \end{pmatrix}.
\end{equation}
In an analogue way, the equation (\ref{generalOscillation}) can get modified. \par 

With these developed tools we can now calculate a general expression for a flavor eigenstate of a neutrino which has mixing with a large number of extra states and also includes mixing with the other flavor states. The investigated case of non-degenerated non-perturbed eigenstates can be used for the ADDM scenario and via a cross check we can reproduce the one-flavor
 equation
 obtained in \cite{Dvali:1999cn}. 
 In the following section we show how one can calculate the flavor states for a highly degenerated mass matrix which are important for the DR scenario.

\section{Highly Symmetric Mass Matrices}

So far we investigated the case of a very general mass matrix which contains mixing with all the states of the extra space. But also the cases where these mass matrices have a specific structure and are highly symmetric are of interest. One specific example for this is the "Many Species Theory" with  exact copies of the SM.

 We now want to present a way how we can deal with this kind of matrices when they are 
  block-wise grouped in their mass matrix. 
  without a loss of generality we illustrate this 
 on example of  the DR scenario.
  A grouping of the different copies of the SM can occur according 
  to the VEV of the Higgs doublets. Notice that 
  even if copies obey a strict permutation symmetry, 
  this symmetry can be spontaneously broken by the 
  VEVs of the Higgs doublets. This is because, due to low cutoff
  and the cross couplings among different doublets, the 
  potential can admit vacua in which 
  Higgs doublets of different copies take different  VEVs, 
  $\langle H_j \rangle = v_j$.
   

Also, because in principal a Majorana mass term for neutrinos is not forbidden neither by gauge nor by permutation symmetry,
we will investigate additionally to the common Dirac operator 
\begin{equation}
(HL)_i \lambda_{ij}\nu_{Rj},
\label{Dirac}
\end{equation}
and also a Weinberg operator of the form
\begin{equation}
(\bar{L^c}i\sigma_2 H)_i \lambda_{ij}(H i \sigma_2 L)_j \,,
\label{Weinberg}
\end{equation}
where the indices $i$ and $j$ label different copies and $L$ beeing the SU(2) doublet and $\sigma_2$ acting in this space.  As previously,  we assume that Yukawa couplings obey the $P(N)$-symmetry 
and therefore have the form of (\ref{Yukawa}). \par 

Notice that the operators (\ref{Weinberg}) break the global lepton 
number symmetries explicitly.

The key now is to assign different Higgs VEVs to
different SM copies.  
We group the copies with the same VEVs in diagonal blocks
of the neutrino mass matrix. 

Let us consider a minimal case of this sort in which 
the VEVs  take two possible values $v$ and $v'$. 
We take a subgroup of size $N < N_{TOTAL}$ and assign the VEV $v$. To the rest of  the species $M = N_{TOTAL} - N$ we assign the VEV $v^{\prime}$.  This assignment can be expressed as,
\begin{equation}
v_i = 
\begin{cases}
v & \text{for } i\leq N\\
v^{\prime} &\text{for } i>N \,.
\end{cases}
\end{equation}
Taking this into account and plugging it into the operators (\ref{Dirac}) and (\ref{Weinberg}) one gets the following mass matrices respectively

\begin{equation}
M^{\text{Majorana}}= \begin{pmatrix}
av^2 & bv^2 & bv^2&... &bv^2& bv^{\prime}v &&...&&bv^{\prime}v 
\\ bv^2 & av^2& bv^2&...&bv^2&\vdots&&\ddots&&\vdots\\

\vdots&&\ddots&&\vdots& \vdots&&\ddots&&\vdots\\

bv^2&&...&&av^2&bv^{\prime}v&&...& &bv^{\prime}v\\

bv^{\prime}v&&...& &bv^{\prime}v&av^{\prime^2}&bv^{\prime^2}&bv^{\prime^2}&...&bv^{\prime^2}\\
 
\vdots&&\ddots&&\vdots&bv^{\prime^2}&av^{\prime^2}&bv^{\prime^2}&...&bv^{\prime^2}\\

\vdots&&\ddots&&\vdots&\vdots&&\ddots&&\vdots\\

bv^{\prime}v&&...& &bv^{\prime}v&bv^{\prime^2}&&...&&av^{\prime^2}
\end{pmatrix},
\label{Majmasmat}
\end{equation}
and
\begin{equation}
M^{\text{Dirac}}= \begin{pmatrix}
av & bv & bv&... &bv& bv &&...&&bv 
\\ bv & av& bv&...&bv&\vdots&&\ddots&&\vdots\\

\vdots&&\ddots&&\vdots& \vdots&&\ddots&&\vdots\\

bv&&...&&av&bv&&...& &bv\\

bv^{\prime}&&...& &bv^{\prime}&av^{\prime}&bv^{\prime}&bv^{\prime}&...&bv^{\prime}\\
 
\vdots&&\ddots&&\vdots&bv^{\prime}&av^{\prime}&bv^{\prime}&...&bv^{\prime}\\

\vdots&&\ddots&&\vdots&\vdots&&\ddots&&\vdots\\

bv^{\prime}&&...& &bv^{\prime}&bv^{\prime}&&...&&av^{\prime}
\end{pmatrix}.
\label{Dirmamat}
\end{equation}
The diagonalization of the above mass matrices will be performed in the next section. 

\subsection{Diagonalizing of the Majorana mass matrices}
In this part, the Majorana mass matrices will be diagonalized. Because the resulting expressions are rather complex the diagonalization procedure will be done within  certain limits. The two limits which will be discussed are $ v^{\prime} \gg v$ and vice versa.


\subsubsection{The symmetric breaking limit of the mass matrix}
Here the focus lies on the Majorana mass matrix (\ref{Majmasmat})
and we make the assumption that the breaking of $P(N)$ is into two equally large sectors, $M=N$. In order to simplify the resulting equations even further, we will also assume that $v^{\prime} \gg v$.
 we put the value $v'$ close to the cutoff of the theory which is $\sim $TeV. This will lead later to very interesting phenomenological implications. \par 

We start diagonalizing (\ref{Majmasmat}) noticing that it is a $2 \times 2$ block matrix.
As the first step, we multiply the matrix with the following transformation matrix
\begin{equation}
U^{\prime} =\begin{pmatrix}
S & 0 \\
0 & S
\end{pmatrix},
\end{equation}
where S is the diagonalization matrix of a matrix of just ones
 (a matrix with the same entry everywhere)
\begin{equation}
S =\begin{pmatrix}
1&-1&\dots&\dots\\
\vdots & 1&0&\dots\\
\vdots& 0&\ddots&0&\dots
\end{pmatrix}.
\end{equation}

This leads then to the following expression

\begin{equation}
U^{\prime -1}M^{Majorana} U^{\prime}= U^{\prime -1} \begin{pmatrix}
A & B \\
C & D
\end{pmatrix} U^{\prime} = \begin{pmatrix}
S^{-1}AS & S^{-1}BS\\
S^{-1}CS & S^{-1}DS
\end{pmatrix},
\end{equation}

where the matrices A, B, D, C denote the block entries of the mass matrix. One can  separate the diagonal entries of the matrices A and D from the rest of the matrix and turn this one into a matrix with just the same entry 
\begin{equation}
A = v\lambda_{ij} = \begin{pmatrix}
(a-b)v^2 &0&\dots\\
0&\ddots&0\\
\vdots&\dots&(a-b)v^2
\end{pmatrix} + \begin{pmatrix}
bv^2 & \dots&\dots\\
\vdots&\ddots&\vdots\\
\vdots&\dots&bv^2
\end{pmatrix}.
\end{equation} 
The diagonal part commutes with $S$ and one is therefore left with the following matrix
\begin{equation}
\begin{pmatrix}
(a-b)v^2 +Nbv^2&0&\dots&Nbvv^{\prime}&0&\dots&0 \\
0& (a-b)v^2&0&\dots&0&\dots&0\\
\vdots&0&\ddots&0&&\dots&0\\
Nbvv^{\prime}&0&\dots&(a-b)v^{\prime^2}+Nbv^{\prime^2}&0&\dots&0\\
0&\dots&&0&(a-b)v^{\prime^2}&\dots&0\\
0&\dots&&\dots&0&\ddots&0
\end{pmatrix}.
\label{1matrix}
\end{equation}

Now one can take out the diagonal element and can bring it down to a $2 \times 2$ matrix of the following form
\begin{equation}
\begin{pmatrix}
Nbv^2 & Nbvv^{\prime}\\
Nbvv^{\prime}& Nbv^{\prime^2} + (a-b)(v^{\prime^2}-v^2)
\end{pmatrix}.
\end{equation}

In order to find the mass eigenstates one has to manipulate \eqref{1matrix} further with the following rotation matrix, 
\begin{equation}
\begin{pmatrix}
cos(\theta)& sin(\theta) \\
-sin(\theta)& cos(\theta) \,
\end{pmatrix},
\end{equation}
with the rotation angel
\begin{equation}
\theta= \frac{1}{2}\arctan(2\frac{vv^{\prime}}{v^{\prime^2}-v^2})\,.
\label{rotsymMaj}
\end{equation}
The rotation matrix multiplied with the $U^{\prime}$ matrix gives the transformation matrix of the mass matrix. The result is

\begin{equation}
U=\begin{pmatrix}
cos(\theta)&-1&\dots&-1&sin(\theta)&0&&\dots\\
cos(\theta)&1&0&\dots&sin(\theta)&0&&\dots\\
\vdots&0&\ddots&&\vdots&0&&\dots\\
-sin(\theta)&0&\dots&0&cos(\theta)&-1&&\dots\\
\vdots&0&\dots&0&\vdots&1&0&\dots\\
\end{pmatrix}.
\end{equation} 
From here  we can see that just two states are affected by the symmetry breaking and the rest stays degenerated with the eigenvalues $(a-b)v^2$ and $(a-b)v^{\prime^2}$. Therefore we can rewrite the new heavy states in terms of the heavy states of the unbroken permutation subset, which we already have encountered in the equations (\ref{eig1}) and (\ref{eigH}). Again in order to simplify the rotation angle (\ref{rotsymMaj}) we use the limit $v^{\prime} \gg v$. The result is then the following
\begin{equation}
n_H^b= n_H - \frac{v}{v^{\prime}} \tilde{n}_H,
\end{equation}
\begin{equation}
\tilde{n}_H^b= \tilde{n}_H+\frac{v}{v^{\prime}}n_H,
\end{equation}
where we used tilde for the $v^{\prime}$ sector.
When one solves now for species states of the two different sectors one gets the following two expressions

\begin{equation}
n_1=\sqrt{\frac{N-1}{N}}n_1^{\prime} + \frac{1}{\sqrt{N}}n_H^b+\frac{1}{\sqrt{N}}\frac{v}{v^{\prime}}\tilde{n}_H^b,
\end{equation}
(notice that for the sake of simplicity the overall normalization factor is suppressed)

\begin{equation}
n_{N+1}=\sqrt{\dfrac{N-1}{N}}\tilde{n}_1^{\prime}+\dfrac{1}{\sqrt{N}}\tilde{n}_H^b-\dfrac{1}{\sqrt{N}}\dfrac{v}{v^{\prime}}n_H^b \,,
\end{equation}
 with the Eigenvalues of the mass eigenstates:
\begin{equation}
m_1^{\prime}=(a-b)v^2,
\end{equation}
\begin{equation}
\tilde{m}_1^{\prime}=(a-b)v^{\prime^2},
\end{equation}
\begin{equation}
m_H= 2(a-b)v^2,
\end{equation}
\begin{equation}
\tilde{m}_H = \text{super massive}.
\end{equation}
This is a rather  interesting result for phenomenology at which we will have a closer look later. We want to point out that the common heavy eigenstate $n_H$ has a mass independent of $N$, which was not the case in the original mechanism. This means that the common heavy eigenstate is not super heavy and neutrino oscillations into this state are therefore possible.

\subsubsection{Asymmetric breaking pattern with a large heavy sector}

 One can also break the symmetry in a way that the sectors include different amounts of copies, $N\neq M$, where 
 $N$ stands for the sector with a VEV of $v$ and $M$ for $v^{\prime}$.  In order to keep the expressions for the final results in a simple form, we take the limit $Mv^{\prime^2} \gg Nv^2$. \\
 After  repeating  the same diagonalization  procedure, the matrix \eqref{1matrix}  in this case has the following form 
\begin{equation}
\begin{pmatrix}
(a-b)v^2 +Nbv^2&0&\dots&Mbvv^{\prime}&0&\dots&0 \\
0& (a-b)v^2&0&\dots&0&\dots&0\\
\vdots&0&\ddots&0&&\dots&0\\
Nbvv^{\prime}&0&\dots&(a-b)v^{\prime^2}+Mbv^{\prime^2}&0&\dots&0\\
0&\dots&&0&(a-b)v^{\prime^2}&\dots&0\\
0&\dots&&\dots&0&\ddots&0
\end{pmatrix}.
\label{Asymmassmatrix}
\end{equation}

Before we can perform the rotation, we have to make an 
intermediate step which brings the off-diagonal entries to the same value. Therefore one applies another transformation matrix of the following form
\begin{equation}
\begin{pmatrix}
1&0&\dots&\dots&\dots&\dots\\
0&\ddots&0&\dots&\dots&\dots\\
\vdots&\dots&\kappa&0&\dots&\dots\\
\vdots&\dots&0&1&0&\dots\\
\vdots&\dots&\dots&0&\ddots&0
\end{pmatrix},
\end{equation}
with $\kappa$ being
\begin{equation}
\kappa= \sqrt{\dfrac{N}{M}}.
\end{equation}

After this procedure the off-diagonal entries are equal and one can perform the rotation like in the symmetric case. 
Correspondingly one gets a mixing angle of the following form
\begin{equation}
\theta=\dfrac{1}{2}\arctan(-2\frac{\sqrt{N}\sqrt{M}bvv^{\prime}}{Nbv^2-Mbv^{\prime^2}}).
\end{equation}

The resulting transformation matrix is 

\begin{equation}
U=
\begin{pmatrix}
cos(\theta)&-1&\dots&-1&sin(\theta)&0&&\dots\\
cos(\theta)&1&0&\dots&sin(\theta)&0&&\dots\\
\vdots&0&\ddots&&\vdots&0&&\dots\\
-\kappa sin(\theta)&0&\dots&0&\kappa cos(\theta)&-1&&\dots\\
\vdots&0&\dots&0&\vdots&1&0&\dots\\
\end{pmatrix}.
\end{equation}
and $\theta$ simplified to
\begin{equation}
\theta = \sqrt{\frac{N}{M}}\frac{v}{v^{\prime}}.
\end{equation}
The resulting mass eigenstates are then 
\begin{equation}
n_H^b=n_H -\frac{N}{M}\dfrac{v}{v^{\prime}}\tilde{n}_H,
\end{equation}
\begin{equation}
\tilde{n}_H^b= \tilde{n}_H + \frac{v}{v^{\prime}}n_H,
\end{equation}
with the eigenvalues
\begin{equation}
m_H= (a-b)v^2,
\end{equation}
\begin{equation}
\tilde{m}_H = \text{super massive}.
\end{equation}

The corresponding copy eigenstates are 

\begin{equation}
n_1 = \sqrt{\frac{N-1}{N}}n_1^{\prime} + \frac{1}{\sqrt{N}} n_H^b + \frac{1}{\sqrt{N}} \frac{N}{M}\frac{v}{v^{\prime}}\tilde{n}_H^b,
\label{phenoW}
\end{equation}
\begin{equation}
n_{N+1}=\sqrt{\frac{M-1}{M}}+\frac{1}{\sqrt{M}}\tilde{n}_H^b-\frac{1}{\sqrt{M}}\frac{v}{v^{\prime}}n_H^b.
\end{equation}

We see that the mass $m_H$ is the same as for the degenerated mass eigenstates.

\subsubsection{Asymmetric breaking pattern with a large light sector}

One can also investigate the case with a large light sector $Nv^2 \gg Mv^{\prime^2}$. In this case the procedure is the same and (\ref{Asymmassmatrix}) stays untouched. The resulting mixing angle is 
\begin{equation}
\theta = - \sqrt{\frac{M}{N}} \frac{v^{\prime}}{v}.
\end{equation}
The eigenvalues are

\begin{equation}
m_H= (a-b)v^{\prime^2},
\end{equation}
\begin{equation}
\tilde{m}_H = \text{super massive}.
\end{equation}
The corresponding eigenstates are given by 
\begin{equation}
\tilde{n}_H^b= \frac{v}{v^{\prime}}n_H + \tilde{n}_H,
\end{equation}
\begin{equation}
n_H^b = \tilde{n}_H - \frac{Mv^{\prime}}{Nv}n_H.
\end{equation}

The copy eigenstates are
\begin{equation}
n_1= \sqrt{\dfrac{N-1}{N}}n^{\prime}_1 - \frac{1}{\sqrt{N}}\frac{v^{\prime}}{v} n_H^b + \frac{1}{\sqrt{N}}\frac{v^{\prime}}{v} \tilde{n}_H^b,
\end{equation}
\begin{equation}
n_{N+1}=\sqrt{\frac{M-1}{M}} \tilde{n}_1^{\prime} + \frac{1}{\sqrt{M}}n_H^b +\frac{1}{\sqrt{M}}\frac{M}{N}(\frac{v^{\prime}}{v})^2 \tilde{n}_H^b.
\end{equation}

Now the situation is  or reversed. The $m_H$ goes to the eigenvalues of the degenerated states of the heavy sector.  Taking $v^{\prime}$ close to the cutoff ($\sim$ TeV) 
the estimated values of $m_H$ could be up to $\sim $keV. 

\subsection{Diagonalizing of the Dirac mass matrix}
Let us now turn to a diagonalization of the Dirac mass matrix which results from the operator (\ref{Dirac}). 
Procedure is similar but 
some details differ from the Majorana case. 

\subsubsection{The symmetric breaking limit of the Dirac mass matrix}
After the first steps similar to the ones taken for the Majorana case, the matrix has the form
\begin{equation}
\begin{pmatrix}
(a-b)v +Nbv&0&\dots&Nbv&0&\dots&0 \\
0& (a-b)v&0&\dots&0&\dots&0\\
\vdots&0&\ddots&0&&\dots&0\\
Nbv^{\prime}&0&\dots&(a-b)v^{\prime}+Nbv^{\prime}&0&\dots&0\\
0&\dots&&0&(a-b)v^{\prime}&\dots&0\\
0&\dots&&\dots&0&\ddots&0
\end{pmatrix}.
\label{2matrix}
\end{equation}
Now the situation is different because the matrix (\ref{2matrix}) is not symmetric (\ref{1matrix}). Because of this one has to introduce the auxiliary parameter $\kappa$ already in the symmetric breaking limit 
\begin{equation}
\kappa = \sqrt{\frac{v^{\prime}}{v}}
\end{equation}
and the rotation angle is
\begin{equation}
\theta= \sqrt{\frac{v}{v^{\prime}}}.
\end{equation}
The resulting heavy eigenstates are 
\begin{equation}
n_H^b = \frac{1}{\sqrt{2}}n_H - \frac{1}{\sqrt{2}}\tilde{n}_H,
\end{equation}
\begin{equation}
\tilde{n}_H^b = \tilde{n}_H +\frac{v}{v^{\prime}}n_H,
\end{equation}
with the eigenvalues
\begin{equation}
m_H = 2(a-b)v,
\end{equation}
\begin{equation}
\tilde{m}_H = \text{super massive}.
\end{equation}

Solving for the the species states leads to 

\begin{equation}
n_1 = \sqrt{\frac{N-1}{N}} n_1^{ \prime} + \sqrt{\frac{2}{N}}n^b_H +\frac{1}{\sqrt{N}}\tilde{n}^b_H,
\end{equation}

\begin{equation}
n_{N+1}= \sqrt{\frac{N-1}{N}}n_1^{\prime} - \sqrt{\frac{2}{N}}\frac{v}{v^{\prime}}n^b_H+\frac{1}{\sqrt{N}}\tilde{n}^b_H.
\end{equation}

\newpage
\subsubsection{Asymmetric breaking pattern with a large heavy sector}
Now we turn again to the cases of asymmetric breaking the permutation group. We investigate the scenario with $M \gg N$. In order to do so,  in the matrix (\ref{2matrix}) we replace
for one sector $N$ with $M$ like in the Majorana case.  In this scenario the auxiliary parameter becomes
\begin{equation}
\kappa= \sqrt{\frac{N v^{\prime}}{M v}},
\label{kappa}
\end{equation}
and the resulting rotation angel is
\begin{equation}
\theta = \sqrt{\frac{Nv}{Mv^{\prime}}}.
\end{equation}
The mass eigenstates are 
\begin{equation}
n_H^b = n_H -\frac{N}{M}\tilde{n}_H,
\end{equation}
\begin{equation}
\tilde{n}_H^b = \frac{v}{\sqrt{v^2 + v^{\prime 2}}}n_H +\frac{v^{\prime}}{\sqrt{v^{\prime 2 + v^2}}}\tilde{n}_H,
\end{equation}
with the eigenvalues
\begin{equation}
m_H = (a-b)v,
\end{equation}
\begin{equation}
\tilde{m}_H = \text{super massive}.
\end{equation}

The species states are 
\begin{equation}
n_1 = \sqrt{\frac{N-1}{N}}n_1^{\prime}+\frac{1}{\sqrt{N}}n^b_H + \frac{1}{\sqrt{N}}\frac{N}{M}\tilde{n}^b_H,
\label{phenoD}
\end{equation}
\begin{equation}
n_{N+1}=\sqrt{\frac{M-1}{M}}\tilde{n}_1^{\prime}+ \frac{1}{\sqrt{M}}\tilde{n}^b_H-\frac{1}{\sqrt{M}}\frac{v}{v^{\prime}}n^b_H.
\end{equation}

Again the oscillation in our copy has an extremely small frequency because the $\Delta m$ goes to $0$ but, on the other hand, it is suppressed as $1/N$ but $N$ in the present case is not large.

\subsubsection{Asymmetric breaking pattern with a large light sector}
Finally, let us  investigate the case with $N\gg M$ and $v^{\prime} \gg v$. The auxiliary parameter $\kappa$ stays the same as in equation (\ref{kappa}). The rotation angel is

\begin{equation}
\theta = - \sqrt{\frac{M v^{\prime}}{N v}},
\end{equation}
with the eigenstates
\begin{equation}
n_H^b= \tilde{n}_H - \frac{M}{N}n_H,
\end{equation}
\begin{equation}
\tilde{n}_H^b = \tilde{n}_H +\frac{v}{v^{\prime}}n_H.
\end{equation}
The eigenvalues are 
\begin{equation}
m_H = (a-b)v^{\prime},
\end{equation}
\begin{equation}
\tilde{m}_H = \text{super massiv}.
\end{equation}
The  corresponding  species states are
\begin{equation}
n_1= \sqrt{\frac{N-1}{N}}n_1^{\prime}-\frac{1}{\sqrt{N}}\frac{v^{\prime}}{v}n^b_H,
\end{equation}
\begin{equation}
n_{N+1}= \sqrt{\frac{M-1}{M}}\tilde{n}_1^{\prime}+\frac{1}{\sqrt{M}}n^b_H.
\end{equation}

\section{Phenomenology}
We now want to turn to the phenomenological implications of the theoretical framework we built up in the  previous  sections. 
We will do this within a specific theory.
  First, we want to point out that the first steps in this topic were already done in \cite{Dvali:1999cn} for ADDM and \cite{Dvali:2009ne} in DR. But in both cases, just one flavour case
of the SM neutrino were investigated. We now aim to generalize this analysis to the three flavour case using the general framework which we presented before. 

\subsection{Phenomenology of ADDM model}

First, we want to discuss the Phenomenology of the ADDM scenario in a realistic three flavour setting. In order to do so, we want to use the framework of section 3 and apply our generally derived formulas to the ADDM case. First we have to define the mass matrix we are investigating. For this we take the Ansatz from \cite{Dvali:1999cn} and generalize it to the three flavour case. To write down the resulting mass matrix we assume that the flavor symmetry is preserved in the bulk. This leads to the effect that the mixing among bulk states is diagonal. The resulting mass matrix is
\begin{equation}
    \begin{pmatrix}
    m_{ee} & \sqrt{2}m_{ee}& \dots &\dots &m_{e\mu}& \sqrt{2}m_{ee}& \dots  &m_{e\tau}&\sqrt{2}m_{ee}& \dots \\
    0 &\frac{1}{R}& 0& \dots&\dots &\dots &\dots &\dots&\dots &\dots\\
    0 &0& \ddots& 0&\dots &\dots &\dots &\dots&\dots&\dots\\
    0 &0&0&\frac{k}{R}&0&\dots &\dots &\dots &\dots&\dots\\
    m_{\mu e} & \sqrt{2}m_{\mu \mu}& \dots&\sqrt{2}m_{\mu \mu} & m_{\mu \mu}& \sqrt{2}m_{\mu \mu}& \dots &m_{\mu \tau}&\sqrt{2}m_{\mu \mu}& \dots \\
    0&\dots&\dots&\dots&0 & \frac{1}{R}&0&\dots&\dots&\dots\\
    \vdots&\dots&\dots&\dots& \dots & 0&\ddots&0&\dots&\dots\\
    m_{\tau e} & \sqrt{2}m_{\tau \tau}& \dots &\dots & m_{e \tau}& \sqrt{2}m_{\tau \tau} &\dots &m_{e\tau}&\sqrt{2}m_{\tau \tau}& \dots \\
    0&\dots&\dots&\dots&\dots&\dots&\dots&0 & \frac{1}{R}&\dots\\
    
    \end{pmatrix}.
\end{equation}
In order to perform the diagonalization of this mass matrix one has to define the parametrization of the $U_{PMNS}$ matrix
\begin{equation}
    U_{PMNS} = \begin{pmatrix}
        c_{12}c_{13}&&c_{13}s_{12}&&s_{13}\\
-c_{23}s_{12} e^{i\phi}-c_{12}s_{13}s_{23}&&c_{12}c_{23}e^{i\phi}-s_{12}s_{13}s_{23}&& c_{13}s_{23}\\s_{23}s_{12}e^{i\phi}-c_{12}c_{23}s_{13}&&-c_{12}s_{23}e^{i\phi}-c_{23}s_{12}s_{13}&&c_{13}c_{23}
    \end{pmatrix}.
\end{equation}
With this PMNS-matrix parametrization we can use the formula (\ref{generalOsc}) to calculate the the expression for 
e.g., the muon neutrino. The result is 


\begin{equation}
    \ket{\nu_{\mu}}= U_{\mu 1}\ket{m_1^e}+U_{\mu 2}\ket{m_1^{\mu}}+U_{\mu 3}\ket{m_1^{\tau}} + \overrightarrow{U_{\mu}}\overrightarrow{C} \sum_{\alpha}\sum_k \frac{1}{k}\ket{k_{\alpha}}.
\end{equation}

with $\overrightarrow{C}_{ADD}$
\begin{equation}
\overrightarrow{C}_{ADD} = \begin{pmatrix}
        \xi^e U_{e1}^{-1}+ \xi^{\mu}U_{e2}^{-1} + \xi^{\tau} U_{e3}^{-1} \\
        \xi^e U_{\mu1}^{-1}+ \xi^{\mu}U_{\mu2}^{-1} + \xi^{\tau} U_{\mu3}^{-1} \\
        \xi^e U_{\tau1}^{-1}+ \xi^{\mu}U_{\tau2}^{-1} + \xi^{\tau} U_{\tau3}^{-1} 
   \end{pmatrix}.
\end{equation}

and the normalisation
\begin{equation}
N^2_e = 1 + \frac{\pi^2}{2}(\overrightarrow{U_{\mu}}\overrightarrow{C})^2.
\end{equation}
 Notice that the parameters $\xi^{\alpha}$ are related with each other via 
 \begin{equation}
     \xi^e \propto m_e \approx \mathcal{O}(1)m_{\mu} \approx \mathcal{O}'(1)m_{\tau},
 \end{equation}
 and therefore the key parameter in this expression is just the size of the dominant extra dimension $R$. \par
 
In order to get an impression on the dependence of this deviation of the composition of an muon neutrino from the Standard Model composition, one can calculate the survival probability. 
 We assume that just the lowest modes of the KK towers contribute to the oscillations, since the higher modes get averaged out due to large mass-splittings. Then the survival probability reads as
 \begin{equation}
     P(\nu_{\mu} \rightarrow \nu_{\mu}) = \frac{1}{|N_{\mu}|^4}\left[\sum_i \sum_j |U_{\mu i}|^2 |U_{\mu j}|^2 e^{\frac{i(m_i^2 - m_j^2)}{2E}} + 3|\overrightarrow{U_{\mu}}\overrightarrow{C}|^4 \left( \frac{\pi^4}{90}-1\right)\right].
 \end{equation}
with $E$ beeing the energy of the investigated neutrino. This can be compared to the original result in \cite{Dvali:1999cn} for the one flavor case
\begin{equation}
    P = \frac{1}{(1+(\pi^2/6)\xi^2)^2}\left[(1+\xi^2)^2 + \left(\frac{\pi^4}{90}-1\right)\xi^4 - \xi^2 \sin^2\frac{(m_n^2 - m_D^2)t}{4E}\right].
\end{equation}
From these two equations one can see that some properties of the one flavor case also appear in a modified way in the three flavor equation. Particularly interesting is how in ADDM models the averaged out modes influence the survival probability by a term proportional to $\left( \frac{\pi^4}{90}-1\right)$ if just the lowest mode is not averaged out. Of course the experimental setup and the specific mass splitting determine how much modes can be resolved in the oscillations. As more modes participate as less important the contribution of the averaged out modes are. \par
\begin{figure}
    \centering
    \includegraphics[scale= 0.5]{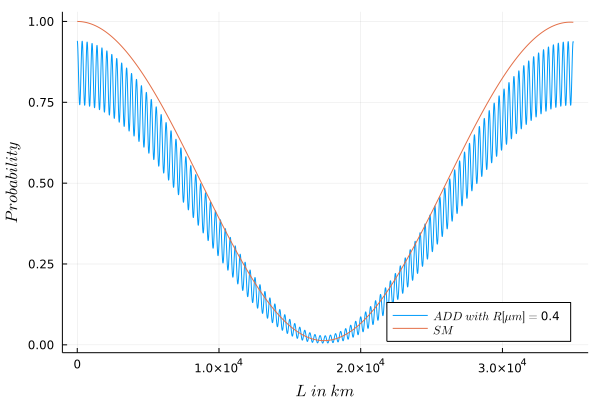}
    \caption{Survival probability of an electron neutrino in a three flavour ADD mixing scenario}
    \label{electronneutrinosurvprobfigure}
\end{figure}

  
For comparizon of the three flavor scenario with SM prediction,  we take the latest results of the nu-fit collaboration \cite{Esteban_2020}  which are: 
\begin{equation}
\theta_{12} = 33,44^{\circ}, \theta_{23}=49,0^{\circ}, \theta_{13}=8,57^{\circ}, \delta_{CP} = 195^{\circ}\,.
\end{equation}
 With this data one can calculate the survival probability of an muon neutrino depending on the parameter $\xi$ of the ADDM model. Figure \ref{electronneutrinosurvprobfigure} shows the result of this calculation and how it deviates from the SM case.
Thus, the precision measurements of neutrino oscillations can put bounds on the critical value $\xi$ of the ADDM model. \par Moreover, this deviation from the SM case also affects the unitarity of the Lepton Mixing Matrix. If the familiar three flavors of the SM 
neutrinos exhaust the spectrum of neutral leptons, 
 the $3 \times 3$ mixing matrix we measure in neutrino experiments must be unitary. This does not hold if there exist more than three neutrinos. In this case the Lepton Mixing Matrix is not a $3 \times 3$ matrix anymore but a $(3+n) \times (3+n)$ matrix where $n$ is the number of additional neutrinos.
 
  Nevertheless, in the experiments that are sensitive to active species,  we would still measure only the $3 \times3$ part of the 
  full Lepton Mixing Matrix.  Since this restricted 
part  in general will not be unitary, we will effectively 
 register a deviation from unitarity. This happens also in the ADDM model where neutrinos can oscillate into the KK modes. Using the above results, the $3 \times 3$ part of the full Lepton Mixing Matrix would get modified in the following way 
\begin{equation}
    \begin{pmatrix}
    U_{ee}\frac{1}{N_e} & U_{e \mu}\frac{1}{N_e}&U_{e \tau}\frac{1}{N_e}\\
U_{\mu e}\frac{1}{N_{\mu}}& U_{\mu \mu}\frac{1}{N_{\mu}}& U_{\mu \tau}\frac{1}{N_{\mu}}\\
U_{\tau e}\frac{1}{N_{\tau}}&U_{\tau \mu }\frac{1}{N_{\tau}}&U_{\tau \tau}\frac{1}{N_{\tau}} \\
    \end{pmatrix}.
\end{equation}
Now the task is to measure the parameters of the well known PMNS matrix very precisely and look for possible deviations from unitarity. Of course, this feature is not unique to ADDM and something similar can be realized in other models too. 
However, the above example provides us with concrete 
motivated framework for setting bounds on the 
unitarity-violating parameters and using them  
for discriminating between the different models. 
We are going to see this explicitly in the next section when 
we discuss phenomenology of the DR model and confront it with ADDM framework. 

%

\subsection{Phenomenology of Dvali-Redi model}
The generalisation of the mass matrix in the DR scenario goes as follows. Of course, the general structure of the mass matrix is again similar to  (\ref{MassMatrix}) but this time the off-diagonal block matrices have the following form 
\begin{equation}
    M_{\alpha \beta} =\begin{pmatrix}
    m_{\alpha \beta} &0 &0 &\dots\\
    0 & m_{\alpha \beta}&0&\dots\\
    \vdots & 0& \ddots&\vdots
    \end{pmatrix}.
\end{equation}

This specific structure comes from the fact that in this theory the mixing among the different flavors can happen within a single  copy since it is determined by the  physics of the SM. This leads to the following electron neutrino eigenstate 
\begin{equation}
    \ket{\nu_e} = \sqrt{\frac{N-1}{N}} (U_{e1} \ket{m_1} + U_{e2} \ket{m_2} + U_{e3} \ket{m_3}) +  \frac{1}{\sqrt{N}} ( U_{e1}\ket{m_1^H} + U_{e2} \ket{m_2^H} + U_{e3} \ket{m_3^H}).
    \label{manyspecies3flavor}
\end{equation}

The key parameter is the number of active species. Above we showed how we can group the total number of species into light and heavy sectors. Now one can investigate different scenarios with sectors which contain different numbers of copies. Because we have access predominantly to our copy of the SM, for us the scenarios with small number of active species in the sector our copy belongs to
are of special interest. Due to this reason we focus on the scenarios with large heavy sectors that bring down the number of active species in our sector. \par 

Taking the two expressions  we found earlier  for the Weinberg- and Dirac operator (\ref{phenoW}) and (\ref{phenoD}) and comparing  them with each other, one sees that the oscillation into the other sector is suppressed by the number of the active species in the large sector. Therefore, one can safely ignore this contribution, especially in the Weinberg case since there it is further suppressed by the scale of the larger VEV $v^{\prime}$. Then the probability of survival in the one-flavour case \cite{Dvali:2009ne} is given by
\begin{equation}
P(t) = 1 - \frac{4}{N}\sin^2( \frac{\Delta m^2 t}{2E}).
\end{equation}
In this scenario, the problem of observing the effect is shifted from the large suppression by the amplitude into the extremely low frequency which comes from very small splitting among the mass eigenstates.  
Nevertheless, this case is still is of high potential 
 interest for long-baseline experiments of neutrino oscillations. 
 Astrophysical sources of high neutrino fluxes could be useful  candidates for testing such scenarios.  
  Of course, detection of deviations from the expected neutrino flux 
  in pure SM requires understanding of the operation 
 mechanisms of these sources to sufficiently high accuracy.  
  

\subsubsection{Integrating out scenario}
We observed that in small light sector scenarios the suppression of the amplitude goes contrary to the frequency of the oscillations. 
 A scenario that can bring both parameters to the range of easier experimental accessibility 
 is the ``integrating out" scenario which we will now consider.
The goal is to combine the advantages of the different 
{above-studied scenarios into an unique setup}. 
 Let us assume that the permutation symmetry is broken very heavily among the two sectors: One sector containing a large number
 $M$ of copies and another sector containing a smaller number $N$. This is the case which we have already discussed above. 
However, let us  now assume that due to additional breaking of the permutation symmetry, the smaller sector is further split into two sectors with the numbers $N^{\prime}$ and $M^{\prime}$.    
Obviously, the primary breaking of perturbation symmetry 
into the $M$ and $N$ sectors is still dominant and the secondary breaking does not affect physics up to effects of order $\mathcal{O}(\frac{N}{M})$ which is already negligibly small. Due to this reason, this sector can be considered as effectively decoupled from the other sectors. 

\par 

Let us now turn our attention to the leftover copies that are broken down into two smaller sectors $N^{\prime}$ and $M^{\prime}$. Here we have a choice to which sector our SM copy belongs. 
%
In particular, we can assume that the number of copies in our sector $N^{\prime}$ is much larger than the other sector $M^{\prime}$. 
This does not decrease the suppression of the amplitude very much but allows us to liberate the value of the common heavy eigenstate $m_H$ in which the neutrinos of both sectors oscillate and can make $\Delta m$ large enough for bringing the frequency to a value comparable to the ordinary oscillations of the SM. 
 This scenario of splitting is analogous to the large light sector scenario. 

\par 

Overall this integrating out scenario enables us to free the both parameters of the theory. It allows us to bring down the number of copies and correspondingly oscillation frequencies 
 to a scale that makes it observable for experiments. \\ 

\begin{figure}
    \centering
    \includegraphics[scale= 0.5]{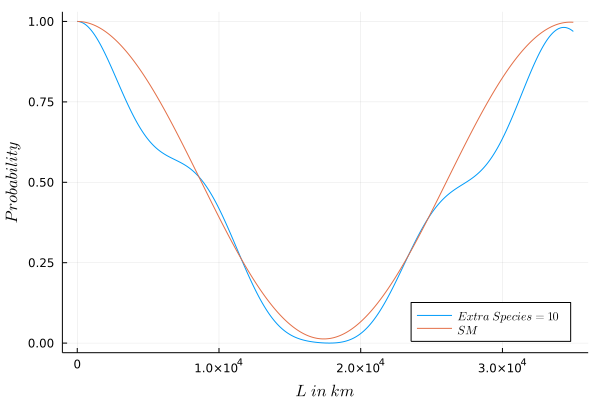}
    \caption{Survival probability of an muon neutrino in a three flavor case in DR model with an equal size splitting scenario}
    \label{electronneutrinosurvprobmanyspeciesfigure}
\end{figure}

We can now calculate the oscillation in the three-flavor case  with an equal size splitting scenario. The equation for the survival probability can be written down as
\begin{equation}
    P(\nu_{\mu} \rightarrow \nu_{\mu}) =\sum_{i=1}^6 \sum_{j=1}^6 |U_{\mu i}|^2 |U_{\mu j}|^2 e^{\frac{i(m_i^2 - m_j^2)}{2E}}.
    \label{largeN}
\end{equation}
First we want to point out that in this expression no modes are averaged out like in the ADDM scenario. The reason for this is because just three additional mass eigenstates have to be included, meanwhile in ADDM scenarios the KK tower can inhabit a very large number of additional mass eigenstates. To analyze equation (\ref{largeN}) further we split it up in the following way
\begin{multline}
     P(\nu_{\mu} \rightarrow \nu_{\mu}) = \left(\frac{N-1}{N}\right)^2 \sum_{i=1}^3 \sum_{j=1}^3 |U_{\mu i}|^2 |U_{\mu j}|^2 e^{\frac{i(m_i^2 - m_j^2)}{2E}} + \frac{N-1}{N^2} \sum_{i=1}^3 \sum_{j=4}^6 |U_{\mu i}|^2 |U_{\mu j}|^2 e^{\frac{i(m_i^2 - m_j^2)}{2E}} \\\\ +\frac{N-1}{N^2} \sum_{i=4}^6 \sum_{j=1}^3 |U_{\mu i}|^2 |U_{\mu j}|^2 e^{\frac{i(m_i^2 - m_j^2)}{2E}} + \frac{1}{N^2}\sum_{i=4}^6 \sum_{j=4}^6 |U_{\mu i}|^2 |U_{\mu j}|^2 e^{\frac{i(m_i^2 - m_j^2)}{2E}}.
\end{multline}
The first term in this expression represents the oscillations within the flavors which are already known in the SM. For large $N$ these oscillations are just slightly modified. One also sees that the dominant contributions are coming from oscillations into the hidden species of order $\frac{1}{N}$ like in the one flavor case in equation (\ref{largeN}). The contributions of solely the BSM terms are suppressed by $\frac{1}{N^2}$.
Figure \ref{electronneutrinosurvprobmanyspeciesfigure} shows the result of the calculations for a muon neutrino.

Let us also show the unitarity violation in the SM lepton mixing matrix 
which is expected by the DR scenario. For this we have to look into the formula (\ref{manyspecies3flavor}).
Picking out the 
$3 \times 3$ block matrix in the upper left corner of the resulting mixing matrix we can write
\begin{equation}
\begin{pmatrix}
\sqrt{\frac{N-1}{N}}U_{ee} & \sqrt{\frac{N-1}{N}} U_{e \mu}& \sqrt{\frac{N-1}{N}} U_{e \tau}\\
\sqrt{\frac{N-1}{N}} U_{\mu e} &\sqrt{\frac{N-1}{N}} U_{\mu \mu}&\sqrt{\frac{N-1}{N}} U_{\mu \tau}\\
\sqrt{\frac{N-1}{N}}U_{\tau e} & \sqrt{\frac{N-1}{N}} U_{\tau \mu} &\sqrt{\frac{N-1}{N}}U_{\tau \tau}
\end{pmatrix}.
\label{manyspeciesunitaritymatrix}
\end{equation}
This is the matrix which is measured by experiments. One can immediately see that unitarity is violated by the overall factor of $\frac{N-1}{N}$. This is a characteristic  signature of the theory and it comes from the democratic oscillation into the 
common heavy eigenstates of the neutrino matrix. The idea of testing the violation of unitarity experimentally by using availible experimental data is a still ongoing work \cite{project}.

\section{Conclusions}
In this paper we have focused on neutrino 
masses  in the class of theories  in which gravity cutoff is lowered 
down to $\sim$TeV scale.  The two main frameworks 
accomplishing this lowering of the cutoff 
 are ADD \cite{ArkaniHamed:1998rs, PhysRevD.59.086004} and  
``many species"  \cite{Dvali:2007hz, Dvali:2007wp} theories.  
In both cases, the  decrease of the gravitational cutoff 
can be understood as a result of ``dilution" of the graviton 
wave function in certain space labelled by a new coordinate.  
In both scenarios the volume of this space can be measured by the number of particle species. Correspondingly the role of the coordinate 
can be played by a species label. As shown in \cite{Dvali:2007hz}, 
in case of ADD \cite{ArkaniHamed:1998rs}
the species 
represent the Kaluza-Klein excitations. Correspondingly, 
the extra space has an actual geometric meaning of large extra space dimensions. 
On the other hand, in ``many species" solution 
to the hierarchy problem \cite{Dvali:2007hz, Dvali:2007wp}, the 
species can be arbitrary particles.  

 Previously it has been suggested that in both scenarios 
the small neutrino masses emerge naturally due 
to the dilution of the wave-function of the sterile (right-handed) neutrino in the extra space.
 Within ADDM this idea was introduced in \cite{Arkani-Hamed:1998wuz} and its phenomenological implications were 
 studied in \cite{Dvali:1999cn}.  
 In this case the wave-function of the sterile neutrino is diluted 
 in the actual geometric extra space.  This results into a highly suppressed Yukawa 
 coupling between the sterile and the active neutrino of the SM, thereby,  
 generating a tiny neutrino mass.   As shown in  \cite{Dvali:1999cn} due to the mixing of 
 active left-handed neutrino with the KK tower of the sterile partner, 
 a non-trivial oscillation pattern emerges.    
 
 More recently, it has been shown  \cite{Dvali:2009ne}
 that a similar suppression 
 mechanism of the neutrino mass works in the DR scenario  \cite{Dvali:2007hz, Dvali:2007wp} in which  
 species represented the identical copies of the SM and the role of  
 the extra coordinate  is played by their label. 
 Using this framework, it was shown in \cite{Dvali:2009ne}
 that the dilution of the wave-function of the sterile neutrino     
 in the space of species results into a small neutrino mass. 
 However,  as discussed there the phenomenological aspects of this scenario are very different from the 
 case of \cite{Arkani-Hamed:1998wuz} which relies on 
 ADDM framework.

In this paper, we have generalized the above original  proposals 
in certain directions.  In particular, we included a more 
realistic case of three SM neutrino flavors. 
 We adopted the universal language of  
 species which allows to capture some general aspects of 
 the neutrino mixing matrix and confront different scenarios.  
   
We calculated an approximate formula for the flavor eigenstates of a general mass matrix using perturbation theory in the three flavor case. Next,  we showed how highly symmetric mass matrices can be calculated in an exact manner and we investigated different symmetry breaking patterns of these highly symmetric mass 
matrices. We gave the explicit expressions for  flavor eigenstates for each case. 

Our further step was to apply this generally derived formulas to the explicit theories of neutrino masses 
such as the proposal of ADDM \cite{Arkani-Hamed:1998wuz, Dvali:1999cn} 
within ADD and the one of DR \cite{Dvali:2009ne} within 
 Many Species frameworks respectively. 
 Here we used the derived formulas and gave a three flavour solution that depends on the parameters of the specific theories.

 As it was already pointed out in \cite{Dvali:1999cn} and \cite{Dvali:2009ne} within ADDM and DR frameworks,  
 the generic prediction of both scenarios is the non-conservation of 
 neutrino number within SM. This is due to the mixing of 
 SM active neutrinos with the tower of sterile partners.  
  This mixing results into the oscillations of neutrinos into 
  hidden species as well as in seeming violation of unitarity 
  within the SM lepton sector.  
 
 Correspondingly, our calculations of these effects for 
 three-flavor case have important 
phenomenological implications in both directions. 
First is the account of deviations of neutrino oscillations from 
the case of SM.  Second, is the parameterization of violation of unitarity of the PMNS-Matrix.

  The structures unitarity-violation in the two different theories
 (ADDM \cite{Arkani-Hamed:1998wuz, Dvali:1999cn} 
of ADD versus Dvali-Redi \cite{Dvali:2009ne} of 
 Many Species)  differ from each other. Our analyses therefore
 has a discriminating power between these two theories.
  
   In general, we can say that 
small neutrino mass generation via mixing with large number 
of extra species is an exciting field with different phenomenological effects on the low energy neutrino physics. These effects can be searched for both in current neutrino experiments, 
such as IceCube \cite{Ahlers2018}, as well as in the planned ones like JUNO \cite{An_2016}. Here, the violation of unitarity can be tested and one can use their results to give bounds on the parameters of the theories such as the size of the extra dimensions in ADDM or the number of sterile neutrino species to which our neutrino mixes within many species scenario.

\section*{Acknowledgments}
I am very thankful to Gia Dvali for his supervision and the advice and comments he offered me during this work. I also would like to thank Goran Senjanovic who taught me a lot about BSM models and neutrino masses in plenty of discussions. This work was partly supported by the IMPRES-EPP and the Sonderforschungsbereich SFB1258.

\bibliography{literatur.bib}

\end{document}